\RequirePackage{fix-cm}
\documentclass[pdflatex,twocolumn,epjc3,final]{svjour3}  
\smartqed  % flush right qed marks, e.g. at end of proof
\RequirePackage{graphicx}
\RequirePackage{mathptmx}      % use Times fonts if available on your TeX system
\RequirePackage[colorlinks,citecolor=blue,urlcolor=blue,linkcolor=blue]{hyperref}
\journalname{Eur. Phys. J. C}
\usepackage{dcolumn}% Align table columns on decimal point
\usepackage{titlesec}
\usepackage{amsmath}
\usepackage{algorithm}
\usepackage{bm}% bold math
\usepackage{multirow}
\usepackage{booktabs}
\usepackage{amsmath,amssymb}
\usepackage{float}

\newcommand{\orcidicon}{\includegraphics[width=8pt]{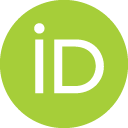}}
\newcommand{\orcidauthor}[1]{%
    \href{https://orcid.org/#1}{\orcidicon}%
}

\begin{document}
\setlength{\parskip}{0pt}
\title{CaloQVAE : Simulating high-energy particle-calorimeter interactions using hybrid quantum-classical generative models}% Force line breaks with \\

\author{Sehmimul Hoque\thanksref{e1,addr1,addr2}\orcidauthor{0009-0003-2992-9200}
        \and
        Hao Jia\thanksref{e2,addr3}\orcidauthor{0000-0002-7391-4423}
        \and
        Abhishek Abhishek\thanksref{addr4,addr3}
        \and
        Mojde Fadaie\thanksref{addr1}
        \and
        J. Quetzalcoatl Toledo-Marín\thanksref{addr4}\orcidauthor{0000-0001-6212-1033}
        \and
        Tiago Vale\thanksref{addr4, addr5}
        \and
        Roger G. Melko\thanksref{e7,addr1,addr2}\orcidauthor{0000-0002-5505-8176}
        \and
        Maximilian Swiatlowski\thanksref{addr4}\orcidauthor{0000-0001-7287-0468}
        \and
        Wojciech T. Fedorko\thanksref{e9,addr4}\orcidauthor{0000-0002-5138-3473}
}

\thankstext{e1}{e-mail: s4hoque@uwaterloo.ca}
\thankstext{e2}{e-mail: haojia@phas.ubc.ca}
% \thankstext{e3}{e-mail: }
% \thankstext{e4}{email: mojde.fadaie@uwaterloo.ca}
% \thankstext{e5}{e-mail: jtoledo@triumf.ca}
% \thankstext{e6}{e-mail: tiago.vale@cern.ch}
\thankstext{e7}{e-mail: rmelko@perimeterinstitute.ca}
% \thankstext{e8}{e-mail: mswiatlowski@triumf.ca}
\thankstext{e9}{e-mail: wfedorko@triumf.ca (corresponding author)}
% You can continue this pattern for the rest of the authors' emails.

\institute{Perimeter Institute for Theoretical Physics, Waterloo, Ontario N2L 2Y5, Canada \label{addr1}
           \and
           Faculty of Mathematics, University of Waterloo, Ontario N2L 3G1, Canada \label{addr2}
           \and
           Department of Physics and Astronomy, University of British Columbia, Vancouver, BC V6T 1Z1, Canada \label{addr3}
           \and
           TRIUMF, Vancouver, BC V6T 2A3, Canada \label{addr4}
           \and
           Department of Physics, Simon Fraser University, Vancouver, BC V5A 1S6, Canada \label{addr5}
           \and
           \emph{Present Address:} University of British Columbia\label{addr3}
}

\maketitle
\begin{abstract}
The Large Hadron Collider's high luminosity era presents major computational challenges in the analysis of collision events. Large amounts of Monte Carlo (MC) simulation will be required to constrain the statistical uncertainties of the simulated datasets below these of the experimental data. Modelling of high-energy particles propagating through the calorimeter section of the detector is the most computationally intensive MC simulation task. We introduce a technique combining recent advancements in generative models and quantum annealing for fast and efficient simulation of high-energy particle-calorimeter interactions.
\end{abstract}
\titlespacing*{\section}
{0pt}{9pt plus 1pt minus 1pt}{4pt plus 1pt minus 1pt}
\titlespacing*{\subsection}
{1pt}{9pt plus 1pt minus 1pt}{4pt plus 1pt minus 1pt}

%\section{Introduction}
\section{Introduction}
The Large Hadron Collider (LHC) is the highest energy particle accelerator in the world, and currently collides protons at $\sqrt{s} = 13.6$~TeV with instantenous luminosity of $2 \times 10^{-34}$ cm$^2$s$^{-1}$.
% Important measurements, such as the discovery of the Higgs Boson in 2012  \cite{ATLAS_discovery,CMS_discovery}, have already occurred, 
Following the discovery of the Higgs Boson in 2012 \cite{ATLAS_discovery,CMS_discovery}, significant advancements in measurements have been achieved, but in order to maximize the physics potential of the collider, an upgrade is underway to significantly increase the instantaneous luminosity to $5-7.5 \times 10^{-34}$ cm$^2$s$^{-1}$. This ``High-Luminosity LHC'' (HL-LHC) dataset will enable significantly more precise measurements of the Higgs boson and other Standard Model particles, and will facilitate searches for new phenomena. 

The significant increase in the rate of data-taking presents a computational challenge in the generation of sufficient high-quality simulated datasets, which are typically used to both calibrate the detectors and evaluate compatibility of the data with different physical hypotheses. These simulated datasets are produced with first-principles particle simulation software in the GEANT4 framework \cite{GEANT4:2002zbu}, though less-accurate, parameterized models of the detectors, often called ``fast simulation'', are also used \cite{ATLAS:2021pzo}. The increased rate of data-taking and precision required to exploit the high-statistics HL-LHC dataset demands both that the computational time to produce simulated datasets decrease, and that the quality of the simulation remains as close as possible to that of the first-principles simulation \cite{CERN-LHCC-2020-015}. %Sehmi - should it be CERN-LHCC-2020-015?

A significant portion of the computational burden of the simulation lies in the interaction of particles with the calorimeter subsystems. Calorimeters are detectors that measure particle energies via the production of showers and the subsequent measurement of those showers as they traverse the active material of the detector. The ATLAS and CMS detectors both contain ``Electromagnetic'' calorimeters dedicated to the measurement of photons and electrons, and ``Hadronic'' calorimeters dedicated to measuring hadrons (typically $\pi^\pm$, protons, and neutrons). The accurate simulation of these complicated particle showers is critical to enable the highest quality measurements, but simulating each shower from first principles is very time consuming. This has motivated in recent years several approaches to reproduce these showers via machine-learning-based methods.

% In this paper, we demonstrate the use of a quantum generative model to generate particle showers at LHC-like detectors. We deploy a restricted Boltzmann machine (RBM) to encode a rich description of particle showers in detectors, and use quantum annealing to avoid computationally expensive Markov Chain evaluations of the shower model. The technique shows promise in both increasing the fidelity of fast simulation, and reducing the computational requirements via a new computing paradigm.

Recently, significant advancements have been made in various machine-learning domains related to generative models. In the development of variational autoencoders (VAEs) and their extensions, a series of breakthroughs have shaped the field \cite{kingma2022autoencoding,rolfe2016discrete,vahdat2018dvae++,khoshaman2018gumbolt}.
% The introduction of auto-encoding variational Bayes in 2014 brought about a stochastic variational inference algorithm that enabled handling of intractable posterior distributions and large datasets \cite{kingma2022autoencoding}. Expanding upon VAEs, discrete variational autoencoders (DVAEs) emerged in 2016, integrating discrete and continuous latent variables through smoothing transformations and hierarchical structures, thus enhancing their representational power \cite{rolfe2016discrete}. Building upon these foundations, overlapping transformations were introduced in 2018 to refine the training of discrete variables in  DVAEs, resulting in state-of-the-art performance \cite{vahdat2018dvae++}. Around the same time, GumBolt was proposed as a model for discrete latent variables with Boltzmann machine priors, leveraging the Gumbel trick to capture complex distributions over discrete variables \cite{khoshaman2018gumbolt}. Collectively, these advancements significantly contribute to the field of VAEs and their extensions, addressing challenges related to intractable posteriors, incorporating discrete variables, and leveraging different priors for enhanced generative modelling.
%Calorimeter Showers:
The development of the generative models provides inspiration to collider physics in the context of calorimeter simulation. In 2018, novel techniques based on generative adversarial networks (GANs) were developed to address the need for fast simulation of electromagnetic showers in calorimeters \cite{CaloGAN}. Later IEA-GAN combines Self-Supervised Learning with GAN to efficiently simulate ultra-high-resolution particle detector responses with a relational reasoning module \cite{IEA-GAN}. In 2021, advancements introduced innovative frameworks like CaloFlow, which are based on normalizing flows and offer fast and accurate simulations with high fidelity and stability \cite{CaloFlow}. In 2022, score-based generative models were introduced for calorimeter shower simulation \cite{PhysRevD.106.092009}. In 2023, CaloFlow v2 further accelerated shower generation by a factor of 500 relative to the original by using probability density distillation \cite{PhysRevD.107.113004}. Most recently, building upon CaloFlow, L2LFlows adapted it to a higher-dimensional setting, thereby improving generative fidelity~\cite{Diefenbacher_2023}.

In deep generative models, the computational costs and the time required for GPU processing are significant concerns. Therefore, we aim to develop a quantum-assisted generative model for calorimeter data simulation, inspired by the potential of quantum computing to reduce these computational burdens. In 2015, a method utilising quantum annealing for training deep neural networks was proposed with a description on how to map Restricted Boltzmann Machines onto a D-Wave Two hardware \cite{Adachi2015}. Building on this idea, a subsequent work explored quantum annealers' ability to sample from Boltzmann distributions \cite{effective_temperatures_estimation}. 
% In 2017, a study tested the freeze-out conjecture in quantum annealers, revealing the quantum fluctuations and noises on freeze-out point \cite{freeze_out}. 
% In 2017, a study on the freeze-out conjecture in quantum annealers highlighted how quantum fluctuations and noises affect the freeze-out point \cite{freeze_out}, motivating us to introduce the real-time temperature estimation to tackle performance discrepancies in quantum devices.
The development of the quantum variational autoencoder (QVAE) showcased the feasibility of utilising a quantum Boltzmann machine as a prior for latent generative process in a variational autoencoder \cite{Khoshaman_2018,winci2020path}.
% These innovations serve to motivate a new research direction: employing quantum generative models for simulating High-Energy Physics (HEP) detectors. 
In 2021, a hybrid quantum-classical qGAN showed promise for accelerating HEP detector simulations using quantum generator circuits and classical discriminator neural networks \cite{qGAN}. In a very recent development, CaloDVAE introduced a discrete variational autoencoder with an RBM prior for fast calorimeter shower simulation, yielding promising results in generating realistic and diverse samples. The possibility of using quantum annealing processors as sampling devices for the RBM prior was also discussed \cite{Abhi}. 

In this paper, we extend the previous work from CaloDVAE to CaloQVAE by employing the D-Wave quantum hardware \cite{DWAVE} to bypass the need for computationally intensive Markov Chain evaluations used in CaloDVAE, and get samples at a high rate from the latent space of a classically trained model.
Our model enables the simulation of showers with clusters that have a granularity and number of cells, as well as energy scales, similar to those expected in actual experiments. This level of detail, corresponding to clusters observed in current and future LHC experiments, marks the first instance of utilizing quantum computing for generative models in calorimeter shower simulations at such a realistic scale.
%\section{Restructured Methodology - Sehmimul}
\section{Model Setup}
% \subsubsection{Architecture}
Variational Autoencoders (VAEs) are a class of latent variable generative models, that maximize an evidence lower bound (ELBO) to the true log-likelihood \cite{kingma2022autoencoding}:\begin{equation}\label{vanilla-vae-loss}
    \text{log }p_{\theta}(\mathbf{x}) \geq \mathbb{E}_{q_{\phi}(\mathbf{z}|\mathbf{x})}[\text{log }p_{\theta}(\mathbf{x}|\mathbf{z})] - \text{KL}[q_{\phi}(\mathbf{z}|\mathbf{x})|| p(\mathbf{z})]
\end{equation}
Equation \ref{vanilla-vae-loss} is the ELBO for the original unconditional VAE framework. $\text{KL}[Q||P]$ is the Kullback-Liebler divergence between two probability distributions, $Q$ and $P$ while $\mathbb{E}_{p}$ denotes an expectation value over the distribution $p$. The variables $\mathbf{x}$ and $\mathbf{z}$ represent data (in this case a vector of calorimeter cell energies) and latent variable respectively. The approximating posterior, $q_{\phi}(\mathbf{z}|\mathbf{x})$ and the generative, $p_{\theta}(\mathbf{x}|\mathbf{z})$ distributions are often parameterized using neural networks with parameters $\phi$ and $\theta$. In the original VAE framework \cite{kingma2022autoencoding}, 
 $p(\textbf{z})$, the prior distribution of the VAE, is the standard Gaussian Distribution $\mathcal{N}(\mathbf{0}, \mathbf{I})$, while $q_{\phi}(\mathbf{z}|\mathbf{x})$ is parameterized by the distribution, $\mathcal{N}(\boldsymbol{\mu}_{\phi}(\mathbf{x}), \boldsymbol{\Sigma}^{2}_{\phi}(\mathbf{x}))$ with mean $\boldsymbol{\mu}_{\phi}(\mathbf{x})$ and covariance $\boldsymbol{\Sigma}^{2}_{\phi}(\mathbf{x})$.

CaloDVAE \cite{rolfe2016discrete} is a hierarchical discrete VAE which extends the VAE framework by introducing discrete latent variables $\mathbf{z}_{i} \in \{0, 1\}$ in the latent space. Equation \ref{vanilla-vae-loss} is the loss for the unconditional VAE model. For conditional VAEs the encoder and decoder have an additional conditioning variable. In our case the auto-regressive encoder, $q_{\phi}(\mathbf{z}|\mathbf{x},\mathbf{e})$, and decoder, $p_{\theta}(\mathbf{x}|\mathbf{z},\mathbf{e})$, are conditioned on energy $\mathbf{e}$ \cite{Abhi,energy-conditioning} (unlike the encoders and decoders of Equation \ref{vanilla-vae-loss} of the unconditional VAE framework) and are modelled by fully connected neural networks. The approximate posterior of the VAE has a hierarchical structure $q_{\phi}(\mathbf{z}|\mathbf{x},\mathbf{e}) = \prod_{i}q_{\phi_{i}}(\mathbf{z}_{i}|\mathbf{z}_{j<i},\mathbf{x},\mathbf{e})$ with $N$ latent groups and the latent distribution of the VAE is modelled by an RBM where $\mathbf{z} = [\mathbf{z}_{1},...,\mathbf{z}_{N}]$ are partitioned into $2$ equal subsets to make the $2$ sides of the RBM \cite{Abhi}.  The RBM is modeled by probability distribution $p_{RBM} = e^{\mathbf{a}_{l}^{T}\mathbf{z}_{l}+\mathbf{a}_{r}^{T}\mathbf{z}_{r}+\mathbf{z}_{l}^{T}\mathbf{W}\mathbf{z}_{r}}/Z$ where $Z$ is the partition function and $(\mathbf{a}_{l},\mathbf{a}_{r},\mathbf{W})$ are the RBM parameters which are trained along with the VAE parameters $(\theta,\phi)$.

At first VAE and RBM parameters are jointly trained using the standard VAE objective and a binary cross-entropy (BCE) loss to enhance the generative capability of the model. For the BCE loss, firstly binary labels are created to isolate pixel positions with non-zero energy values in the energy images. Subsequently, a BCE loss is computed between the binary label of the preprocessed input data and the binary label of the reconstructed data to incentivize the VAE to faithfully reproduce the distribution of non-zero energy pixels \cite{Abhi}.
Furthermore, due to non-differentiability of discrete values we replace discrete values with continuous variables following the Gumbel trick to allow backpropagation during training \cite{khoshaman2018gumbolt}. Therefore the latent discrete variables $\mathbf{z}$  are replaced with continuous proxy variables $\zeta$ with the Gumbel trick. Similarly we also apply Gumbel trick on reconstructed labels to mask the reconstructed energies. Finally in inference the continuous variables are replaced with the discrete variables \cite{Abhi}.

We obtain a trained model with trained VAE and RBM parameters using the prescription above. The next step is to generate latent RBM samples, $\mathbf{z}$, and pass the samples through the trained VAE decoder to obtain the shower images. In CaloDVAE, block Gibbs sampling is used to generate the latent RBM samples. In CaloQVAE, we propose a novel calorimeter simulation framework based on the quantum variational autoencoder generative model to generate latent space samples, instead of using computationally expensive block Gibbs sampling.
\section{CaloQVAE Training and Inference}

In this work we demonstrate that we can generate latent space samples using the D-Wave 2000Q annealer. Note that Quantum Annealers are mostly known to be used to find ground states of Hamiltonians to which some optimization problems can be translated to~\cite{10.1145/2482767.2482797}. However, due to interactions with the environment there is some probability that at the end of anneal, the final system will be in an excited state instead of the ground state. The probability distribution of the excited states follow the Boltzmann distribution, and hence D-Wave can also be used as an effective sampler from that distribution~\cite{Adachi2015}.  

The first step to use the quantum processing unit (QPU) in the pipeline is to embed the RBM on the quantum hardware. However, the QPU architecture is not fully connected and instead follows a Chimera graph topology, making it impractical to natively embed a fully connected RBM in the QPU~\cite{Adachi2015}. We therefore create a masking function to remove selected edges from an RBM to produce a Chimera RBM. We train the model classically analogously to the CaloDVAE training by replacing the RBM of CaloDVAE with a Chimera RBM.

In the inference phase, we first convert the Chimera RBM Hamiltonian, $\mathcal{H}_{\text{P}}$, to an Ising model Hamiltonian using a linear transformation \cite{Dixit2022}. We then embed the Chimera RBM on the QPU and anneal according to the QPU Hamiltonian $\mathcal{H}(s) = A(s)\mathcal{H}_{\text{T}} + B(s)\mathcal{H}_{\text{P}}$, where $\mathcal{H}(s)$ is the total Hamiltonian of the system at the rescaled time $s$ during annealing, $\mathcal{H}_{\text{T}}$ is the transverse field Hamiltonian, and $A(s)$ and $B(s)$ are monotonic functions \cite{amin2018quantum}. The annealing process follows the standard D-Wave annealing schedule and lasts for $20 \ \mu\text{s}$. During this process, a freezing point $s^{*}$ is typically observed on the device \cite{AminFreeze,Marshall2017}, after which $A(s^{*}) \ll B(s^{*})$ and the system's dynamics change negligibly. If the freezing time is sufficiently short, the resulting configurations $x$ from the device will approximately follow a Boltzmann distribution at $s^{*}$, given by $P(x) \propto \exp{(-\beta_{\text{eff}}^{*}\mathcal{H}_{\text{P}}(x))}$. This defines an effective inverse temperature $\beta_{\text{eff}}^{*}$ \cite{Adachi2015,Perdomo2015}, which varies with different programming cycles and depends on several unknown parameters \cite{freeze_out}. Our goal is to generate samples from the distribution $P_{\text{ideal}}(x) \propto \exp{(-\mathcal{H}_{\text{P}}(x))}$, as $P_{\text{ideal}}(x)$ represents the distribution of the Chimera RBM from which we wish to sample. To achieve this, if $\beta_{\text{eff}}^{*}$ was known, the model parameters of $\mathcal{H}_{\text{P}}(x)$ could be scaled by $\beta_{\text{eff}}^{*}$ to match the desired distribution $P_{\text{ideal}}(x)$. Therefore, since $\beta_{\text{eff}}^{*}$ is not known a priori and varies with different instances, real-time estimation of $\beta_{\text{eff}}^{*}$ is necessary \cite{XuAdaptive2021}. A single run of this scaling parameter estimation is needed, after which numerous samples from the QPU can be rapidly obtained. 
% In our work we estimate $\beta_{eff}^{*}$ using an iterative procedure~\cite{Winci_2020} where we first generate samples from the trained Chimera RBM classically using Block Gibbs Sampling. We then iteratively generate synthetic samples from the QPU while adjusting $\beta_{eff}^{*}$ and then rescaling the QPU parameters (couplings and local fields) by $\beta_{eff}^{*}$ until the quantum energy distribution converges to the classical energy distribution~\cite{Winci_2020}. Figure~\ref{fig:betahist} shows energy distributions for the QPU samples and classical samples.

\begin{figure}[h]
%\begin{center}
\includegraphics[width=\linewidth]{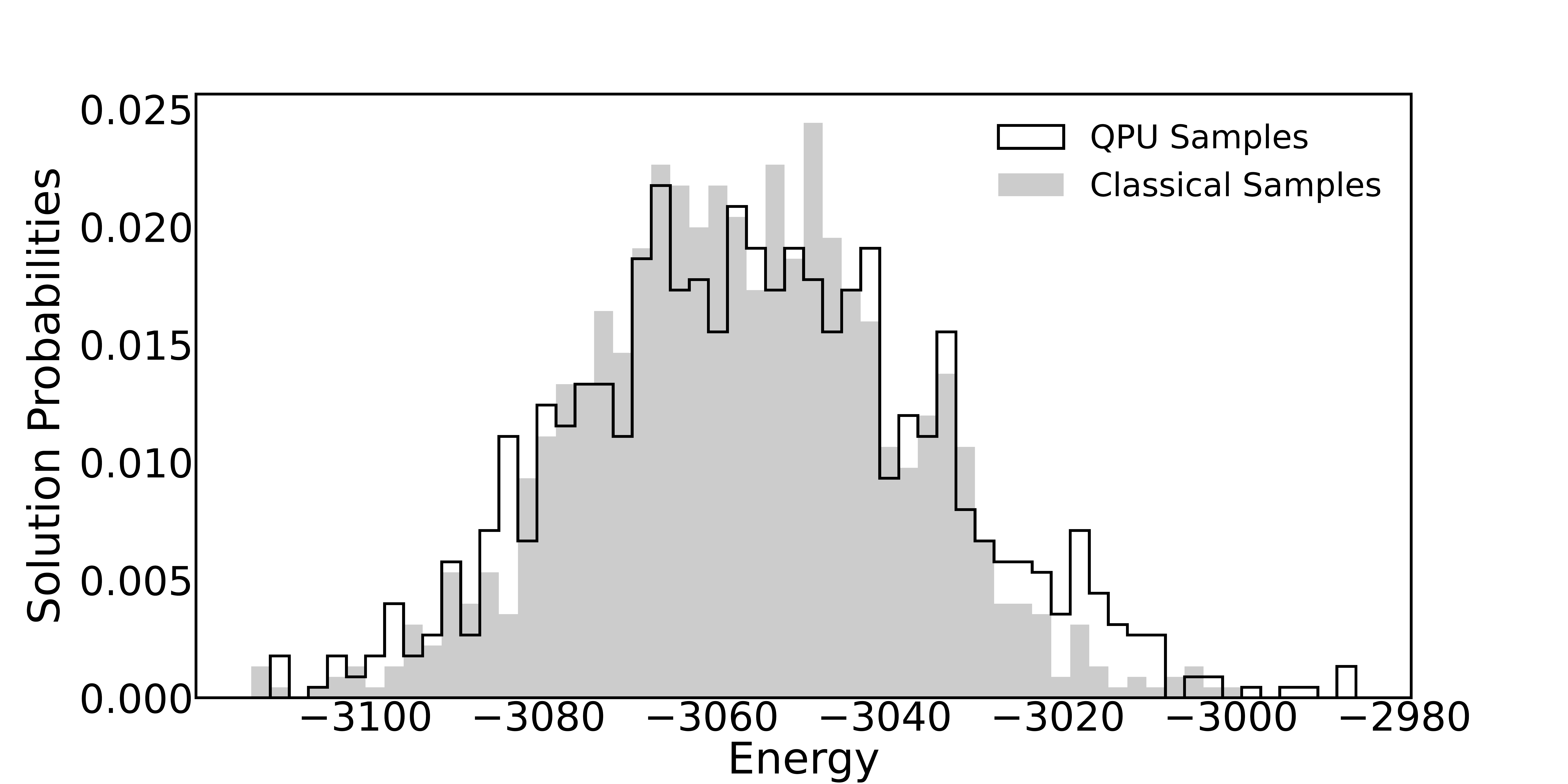}
%\end{center}
% \captionsetup{labelformat=empty, justification = raggedright}
% \captionsetup{justification=raggedright,singlelinecheck=false}
\caption{Histogram showing the probability of obtaining sample $s$ with Ising Energy computed by $E(s) = s^{T}Js+h^{T}s$ in dimensionless units for a trained CaloQVAE model. The energy distribution of QPU samples (solid line) are very close to the classical samples (shaded) after accounting for the $\beta_{eff}^{*}$ factor.} 
\label{fig:betahist}
\end{figure}

In our work, we estimate $\beta_{\text{eff}}^{*}$ using an iterative procedure \cite{winci2020path}. The process begins with the classical generation of samples from the trained Chimera RBM using block Gibbs sampling. Then synthetic samples are iteratively generated from the QPU, and $\beta_{\text{eff}}^{*}$ is updated following an update rule~\cite{winci2020path} to align the mean energies of the RBM and D-Wave samples. This involves rescaling the QPU parameters (couplings and local fields) by the updated $\beta_{\text{eff}}^{*}$ for each iteration. The $\beta_{\text{eff}}^{*}$ training process is repeated over multiple ensambles, with close monitoring of the $\beta_{\text{eff}}^{*}$ values until they converge to a stable value. Figure~\ref{fig:betahist} illustrates the energy distributions for both QPU and classical samples after the procedure concluded, demonstrating convergence of the quantum annealing distribution to the one obtained from the classcal RBM. In CaloQVAE, with a trained model, the $\beta_{\text{eff}}^{*}$ estimation is performed only at the start of the sampling process to determine the current temperature and does not need to be repeated for subsequent D-Wave sampling requests.

\section{Performance Evaluation}

\begin{figure*}
% \begin{center}
\includegraphics[width=1\linewidth]{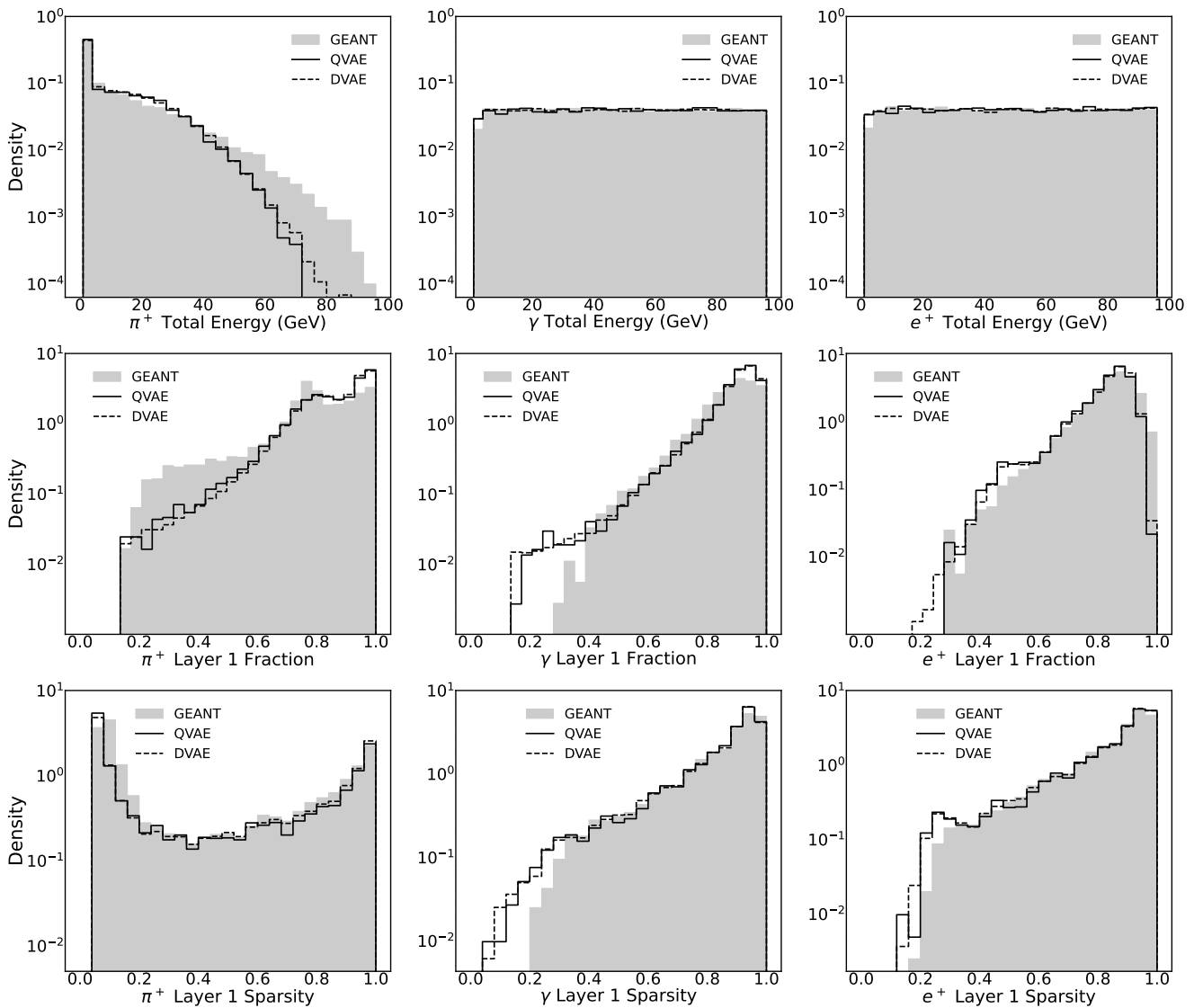}
% \end{center}
% \caption{Comparison of selected shower shape variables for positrons (left sub-panel pair) and charged pions (right sub-panel pair) in a calorimeter. Each pair of sub-panels shows the distributions of two key variables: (1) Fraction of energy: The energy deposited in the first layer compared to the sum of energy in all layers. (2) Sparsity: The ratio of the number of hits in the first layer to the total number of hits in all layers. Each variable is plotted for both particle types to illustrate the close alignment among the results from GEANT4, DVAE (classical) and QVAE (quantum) models.} 
\caption{Comparison of selected shower shape variables for three clusters ($\pi^+$, $\gamma$, and $e^+$). The rows of sub-panels show the distributions of three key variables: (1) Total Energy: The total energy observed across all layers. (2) Energy Fraction: The fraction of energy deposited in the layer 1 (middle layer) compared to the total energy in all layers. (3) Sparsity: The ratio of the number of hits in the layer 1 to the total number of hits across all layers. Each sub-panel illustrates the close alignment of results from the GEANT4, DVAE (classical), and QVAE (quantum) models.} 
\label{fig:allhist}
\end{figure*}

For training and performance evaluation we use the public CaloGAN dataset \cite{Nachman2017} simulated from GEANT4 \cite{GEANT4:2002zbu}. The dataset provides 100,000 events for $\pi^{+}$, $\gamma$ and $e^{+}$ incident particles~\cite{CaloGAN} and is split into 80\%, 10\%, and 10\% subsets for training, validation, and testing, respectively. After conducting a hyperparameter scan \cite{Abhi}, we obtain the best models for each particle type. Utilizing classical block Gibbs sampling and QPU sampling methods respectively, we generate new samples from the latent space and pass them through the decoder to obtain synthetic shower images. These images are then compared with GEANT4 test data to assess their performance. Firstly, chi-square and KS tests were conducted on the total event energies for CaloDVAE and CaloQVAE samples in the 1-100 GeV incident energy range. The results indicated consistency between the two sets of synthetic samples, further supporting the successful utilization of the QPU for sampling from the classically trained latent space. Compared with GEANT4 samples on chi-square tests, CaloQVAE performs similarly to the ones already adopted, within an order of magnitude \cite{ATLAS:2021pzo}. Moreover, classification can be used as a useful indicator of the performance of generative models. A good classifier should be able to understand and capture the important patterns in the data because it needs to learn discriminating features that distinguish different classes or categories accurately. We perform binary classification of $e^{+}$ vs $\pi^{+}$ and $e^{+}$ vs $\gamma$ using a neural network with 6 fully connected layers with dropout, following the method of \cite{CaloGAN}. All synthetic data is generated from the QPU using separate models trained separately on $e^{+},\pi^{+}$ and $\gamma$ samples. Our accuracy metrics (Table \ref{tab:accuracy}) are very close to the results obtained in \cite{CaloGAN}. The similar accuracy metrics illustrate that our CaloQVAE model is generating synthetic data that captures important patterns present in the true data, which indicates that the synthetic images generated from the QPU can represent the underlying structure and characteristics of the true data. The fact that the classifier trained on synthetic $e^{+}$ vs $\pi^{+}$ data can perform well when tested on the GEANT4 data, and vice versa, suggests that the features learned by the CaloQVAE model are transferable. This means that the CaloQVAE model can encode the relevant information from the true data into its latent space representation, which allows the generated data to be informative and useful for training a classifier. An issue that requires attention is the unexpected high performance of the $e^{+}$ vs $\gamma$ classifier, which is trained with synthetic data and tested on a separate set of synthetic data. To address the issue, we checked the datasets and found that the mean energy distributions in 504 calorimeter cells were consistent for $e^{+}$ and $\gamma$ synthetic datasets. We tried solutions like masking low-energy pixels and unifying dataset normalization, but the issue of over-performance persists. We believe that such performance difference is due to an existing intrinsic bias in the GEANT4 dataset between $e^{+}$ vs $\gamma$. As shown in Table \ref{tab:accuracy}, training and testing both on GEANT4 data for positron vs gamma already yield a relatively high classification rate compared to expectations. This rate increases further when training on GEANT4 and testing on synthetic data. Our model learns and amplifies this bias from the training dataset, leading to the observed issue. Similar over-performance has also been observed in \cite{CaloGAN}.
% This raises the possibility that our CaloQVAE model may generate certain non-physical features, which allow the classifier to distinguish between the $e^{+}$ and $\gamma$ samples. Further research is warranted to explore potential ways of addressing this challenge.
\begin{table}
\caption{Mean (standard deviation) of the binary classification accuracy over 10 particle classification trials for $e^{+}$ vs $\pi^{+}$ and $e^{+}$ vs $\gamma$ where we train and test both on the GEANT4 data and QPU synthetic data.}
\label{tab:accuracy} 
\resizebox{240pt}{!}
{
\begin{tabular}{ c c c c c c } 
\multicolumn{2}{c}{} & \multicolumn{2}{c|}{$e^{+}$ vs $\pi^{+}$} & \multicolumn{2}{c}{$e^{+}$ vs $\gamma$}\\ [0ex] 
\multicolumn{4}{c|}{} & \multicolumn{2}{c}{}\\[\dimexpr-\normalbaselineskip]
\hline
\multicolumn{2}{c}{} & \multicolumn{2}{c|}{Test} & \multicolumn{2}{c}{Test}\\ [1ex]
\multicolumn{2}{c}{} & \multicolumn{1}{c}{Geant4} & \multicolumn{1}{c|}{Synthetic} & \multicolumn{1}{c}{Geant4} & \multicolumn{1}{c}{Synthetic} \\[\dimexpr-\normalbaselineskip+3ex]
\multirow{2}{*}{Train} & \multicolumn{1}{c}{Geant4} & \multicolumn{1}{c}{99.8 ($<$0.1)} & \multicolumn{1}{c|}{99.8 ($<$0.1)} & \multicolumn{1}{c}{67.1 (0.5)} & \multicolumn{1}{c}{74.8 (2.8)}\\
& \multicolumn{1}{c}{Synthetic} & \multicolumn{1}{c}{90.6 (4.0)} & \multicolumn{1}{c|}{100.0 ($<$0.1)} & \multicolumn{1 }{c}{53.0 (1.0)} & \multicolumn{1}{c}{99.9 ($<$0.1)}\\
\end{tabular}
}
\end{table}

We qualitatively analyze one-dimensional histograms of the typical shower shape variables. As shown in Fig \ref{fig:allhist}, there is a close alignment between the results from GEANT4 data and generative models, which signifies that our generative model effectively captures the underlying patterns and distributions of the data. The consistent distribution patterns highlight the generative model's fidelity in representing the essential features in the training dataset. Despite the observation of tail differences at the low fraction region for the positron in layer 1's fraction histogram, the overall similarity between the performance of QVAE and DVAE models indicates our pipeline for $\beta_{\text{eff}}^{*}$ estimation and scaling is well behaved so that the quality of samples from the quantum device is on par with those from the block Gibbs method.

\begin{figure}
\centering
\includegraphics[width=0.9\linewidth]{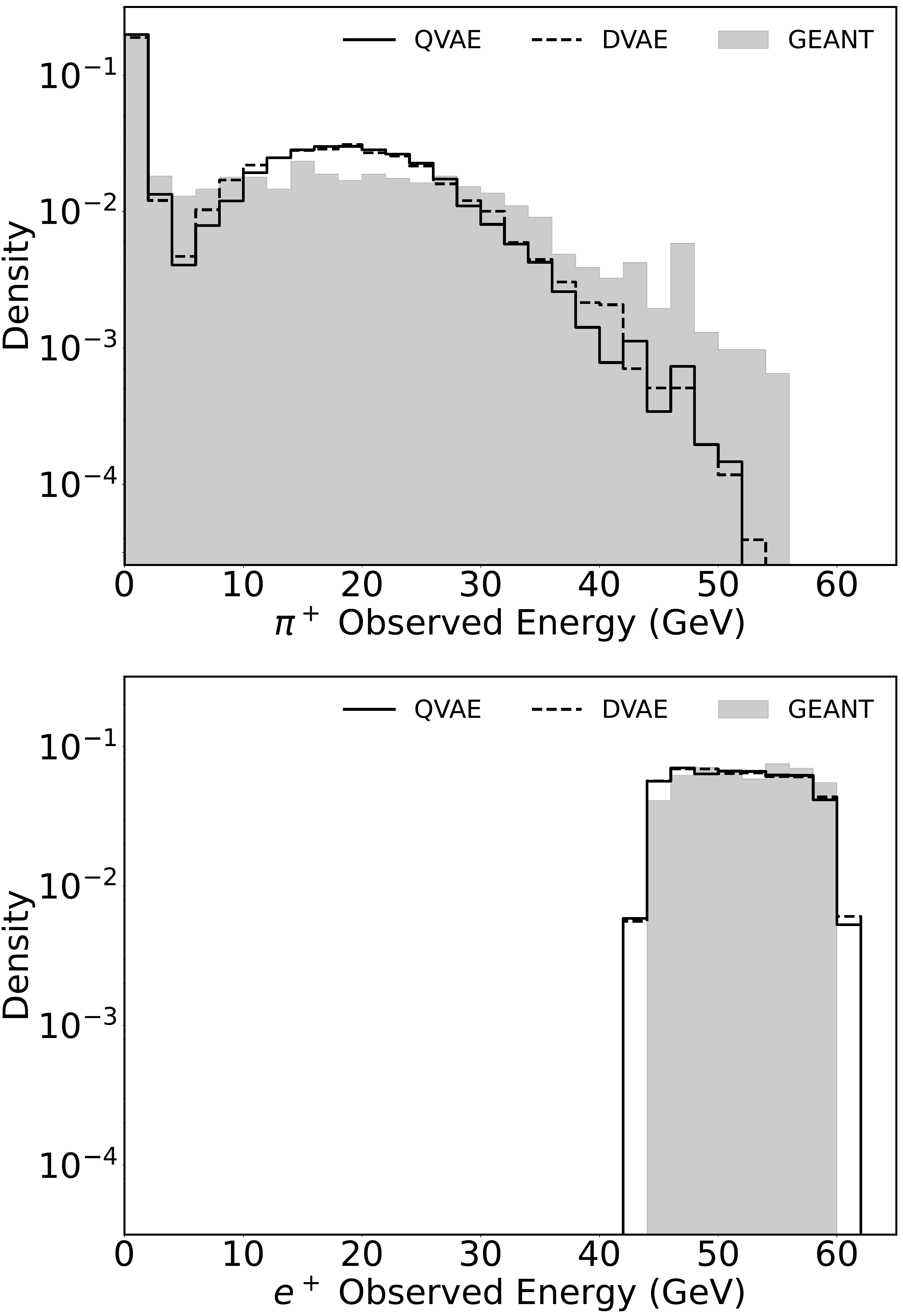}
\caption{Comparison of energy spectra between CaloDVAE (classical) and CaloQVAE (quantum) models against GEANT4 test data for incident particle energies ranging from 45 to 60 GeV. The data encompasses both pion (top panel) and positron (bottom panel) clusters.} 
\label{fig:sliced_energy}
\end{figure}

To assess the practical deployment performance of the model's energy conditioning, we generate samples from the trained model requesting a specific range of incident energy and histogram the observed energy. We generate the samples by classical and quantum approaches and compare the results with a GEANT4 test dataset under the same range of incident energy. As shown in Fig \ref{fig:sliced_energy}, for the positron cluster, the conditioning response of DVAE or QVAE can form a flat plateau with good accuracy in reproducing the defined selection of the input energy range. The pion cluster has a broadened response that is due to the nature of the unconstrained charged pion shower in the electromagnetic calorimeter. The shower shape histograms for both electron and pion clusters exhibit a noticeable similarity to the characteristics seen in the true data. This agreement indicates that our model has capability to capture and reproduce patterns from the first-principles simulation, suggesting its potential as a useful tool for calorimeter shower simulation.

% The histograms of the generative models for both electron and pion clusters are able to mirror the characteristics observed in the true data. This compelling agreement underscores the VAE's ability to capture and faithfully reproduce the intricate patterns inherent in the first-principles simulation, thereby affirming its efficacy as a powerful tool for energy pattern generation. 

\par The monitoring results of the DVAE sampling rate show efficient performance on our NVIDIA GeForce RTX 2080 Ti with 11264 MB memory. The processing time per sample for latent space sampling takes the most substantial portion ( $\sim$ 0.5
$\mathrm{ms}$), followed by data loading to the GPU ( $\sim$ 10 $\mathrm{ns}$), and passing through the decoder ($\sim$ 1 $\mathrm{\mu s}$). The DVAE model exhibits similar sampling rate with the CaloGAN results \cite{CaloGAN} up to an order of magnitude. On the other hand, the quantum sampler achieves slightly faster sampling rate at 0.4 $\mathrm{ms}$ per sample, but impressively, the core QPU annealing rate is only 20 $\mathrm{\mu s}$ per sample, hinting at the raw speed of quantum processing. The extra QPU access time is mainly used for tasks like readout and handling delays, which have not been thoroughly optimized yet because these overheads are a much smaller fraction for more traditional quantum annealing applications. This suggests that with optimized engineering, the readout time could be optimized and quantum sampling could vastly outpace classical methods in efficiency for generative modeling, making it well-suited for real-time and large-scale applications.
% The extra QPU access time is mainly used for tasks like readout and handling delays, suggesting that with optimized engineering, quantum sampling could vastly outpace classical methods in efficiency for generative modeling, making it well-suited for real-time and large-scale applications. 

% We expect to apply the QVAE model to more complex objects, jets and events in the future, for which annealing time is expected to remain at the current level.

\section{Conclusion}
This work is the first demonstration of the application of a quantum annealing device to the computationally expensive simulation of particle showers at the Large Hadron Collider. We have shown that it is possible to utilize the Quantum Processing Unit (D-Wave Chimera 2000Q) to generate RBM samples which can be used to generate
particle showers. The QPU-generated events show good alignment with the classically Monte Carlo-generated events. While the present engineering challenges may not be fully mitigated in the future, the raw QPU annealing rate of 20 $\mu s$ hints at a potential order of magnitude speed increase over GPU sample generation. 
% While further work remains in evaluating on more realistic simulation and a wider range of physics topologies, the initial results demonstrate CaloQVAE to be a promising application of quantum computing to open research questions in fundamental physics.
The initial results demonstrate that CaloQVAE is a promising application of quantum computing to open research questions in fundamental physics. However, further work is needed to evaluate its performance on more realistic simulations, such as the CaloChallenge datasets \cite{michele_faucci_giannelli_2023_8099322,faucci_giannelli_2022_6366271,faucci_giannelli_2022_6366324} and across a wider range of physics topologies.

%This work is the first demonstration of quantum application for a problem at the appropriate scale for the LHC to our knowledge. We have successfully implemented a Variational Autoencoder model with a chimera RBM prior where samples from D-Wave 2000Q QPU were used in inference. In future work, this work can be extended to use QPU samples during training as well. Furthermore, a more detailed and theoretical study can be done to investigate if there is any quantum advantage if we sample from a QPU. 

\begin{acknowledgement}The authors would like to thank Drs. Walter Vinci, Mohammad Amin, Geoffrey Fox, the members of the DeGeSim Collaboration as well as the members of the Perimeter Institute Quantum Intelligence Lab for many enlightening discussions. We thank Dr O. Di Matteo for early project conceptualization and material support for A. Abhishek during the final stages of the research. This work is supported by the Natural Sciences and Engineering Research Council of Canada (NSERC), National Research Council (NRC) and the Canadian Institute for Advanced Research (CIFAR) AI chair program. 
Research at Perimeter Institute is supported in part by the Government of Canada through the Department of Innovation, Science and Economic Development Canada and by the Province of Ontario through the Ministry of Economic Development, Job Creation and Trade. 
We gratefully acknowledge funding from the NRC via Agreement AQC-002, NSERC, including via grants SAPPJ-2020-00032, SAPPJ-2022-00020, and RGPIN-2022-04609
This research was enabled in part by computational support provided by the Shared Hierarchical Academic Research Computing Network (SHARCNET) and the Digital Research Alliance of Canada.
\par S.H. and H.J. contributed equally to this work and are listed in alphabetical order.
\\\\\textbf{Data Availability Statement} The analytical results and code are available online \cite{githubQVAE}.
\end{acknowledgement}

% \appendix

% \bibliographystyle{spbasic}      % basic style, author-year citations
% \bibliographystyle{spmpsci}      % mathematics and physical sciences
\bibliographystyle{spphys}       % APS-like style for physics
\bibliography{caloqvae_paper}% Produces the bibliography via BibTeX.
\end{document}